\documentclass{egpubl}  

\ifpdf \usepackage[pdftex]{graphicx} \pdfcompresslevel=9
\else \usepackage[dvips]{graphicx} \fi

\PrintedOrElectronic

\usepackage{t1enc,dfadobe}

\usepackage{egweblnk}
\usepackage{cite}

\newenvironment{tightList}
{\begin{list}{}{\partopsep=\baselineskip
	\parskip=0pt
	\parsep=0pt
	\topsep=0pt
	\itemsep=0pt
	\labelwidth=0pt
	\itemindent=-10pt}}
{\end{list}}

\usepackage{amsmath,amsthm,amsfonts,amssymb,verbatim, stmaryrd} 

\usepackage{float}
\usepackage{graphicx}

\usepackage{url}
\usepackage{hyperref}

\title{Retinal Vessel Segmentation Using A New Topological Method}
\author{Martin.Brooks@varilets.org}

\begin{document}       

\maketitle

\begin{abstract}
A novel topological segmentation of retinal images represents blood vessels as connected regions in the continuous image plane, having shape-related  analytic and geometric properties. 
This paper presents topological segmentation results from the DRIVE retinal image database \cite{staal}.
\end{abstract}


\section{Introduction}

Retinal image analysis aids disease detection and diagnosis \cite{Abramoff, Bhaduri, Jelinek, Sreejini}, with segmentation of retinal vessels as an important subproblem.
Medical image segmentation is an active research area (e.g.\ survey articles \cite{Zanaty, Ghnomiey, Zhao}) having specialized techniques for retinal vessel segmentation, including mathematical transforms \cite{Selvathi, Soares, Wang, Fazli, Cummings2010, Sadeghzadeh},  signal processing and statistics \cite{Frangi1998, Singh, Oliveira, Sumathi, GeethaRamani, Raja}, level sets \cite{QianZhao, Gongt, Dizdaroglu2014, Lathen}, fractal characteristics \cite{Stosic, Paripurana, Huang}, deep neural networks \cite{Fang, Melinscak, Maji}, as well as region, contour and many other methods described in survey articles (e.g.\  \cite{Fraz, Gupta}).

This paper introduces a novel \emph{topological} image segmentation technique\footnote{The retinal vessel literature typically uses ``topology'' in reference to vessel branching structure, however in this paper it refers to the \emph{analytic topology} \cite{Whyburn1942} of an image's continuous interpolation.}. Unique features include:

\begin{tightList}
\item[(1)] Transformation from raster graphics to scalable vector graphics (SVG) \cite{Eisenberg, HuangSVG, Mohammed} in the continuous image plane. 
\item[(2)] Representation of each image segment as a connected vector graphics shape, having boundary comprised of one or more simple closed curves, each represented as a circular sequence of hyperbolic or linear segments. 
\end{tightList}

Vector graphics in real coordinates (\#1) means that image segmentations can be closely inspected by enlargement. 
Analytic representation (\#2) allows for quantitative analysis and pattern recognition among the segments in order to recognize and measure vessels.

We demonstrate that \emph{large} vessels, i.e.\ relative to image resolution, are mostly comprised of image segments having identifiable visual and analytic characteristics.
Often, an individual segment corresponds to a large connected portion of vessel branching structure.

We use the topological segmentation method of \emph{varilet image analysis} \cite{varilet_image}. Roughly, varilets may be thought of as a topological analogue to wavelets.
Retinal vessel segmentation is well suited to varilet analysis.

We have used the DRIVE \emph{training set} to determine all varilet analysis parameters; we then used these parameters for  segmentation of the DRIVE \emph{test set}.

Varilet image segmentation is hierarchical; each segment is recursively segmented until the finest level of detail is reached. For retinal vessel segmentation we use only the two coarsest levels of the hierarchy. 

Application of varilet segmentation was fully automatic and unsupervised, with the green channel of a DRIVE jpeg image\footnote{Varilet analysis is contrast-independent. The only preprocessing \cite{Rasta} was to ensure that all pixels of the black background are truly black.} as input to Common Lisp code that generates SVG images.

The SVG images provide subjective evidence that topological image segmentation may be a useful preprocessing step for retinal vessel segmentation. 
The text of figures \ref{fig_3} \& \ref{fig_4} suggests how preprocessing by topological segmentation may be followed by pattern recognition.
Section \ref{int_sec} suggests how pixel-based methods, e.g.\ incorporating knowledge of the retina, may be integrated with topological segmentation.

\section{Example}

The images in this section are jpeg renderings of scalable vector graphics, and can be viewed only with moderate magnification.
For access to high-resolution jpeg and SVG images, see  \url{http://varilets.org}.

The SVG images were enhanced by a 3D effect and black segment boundaries, to emphasize the individual segments.

\begin{figure}[H]\centering
\includegraphics[width=3.3in]{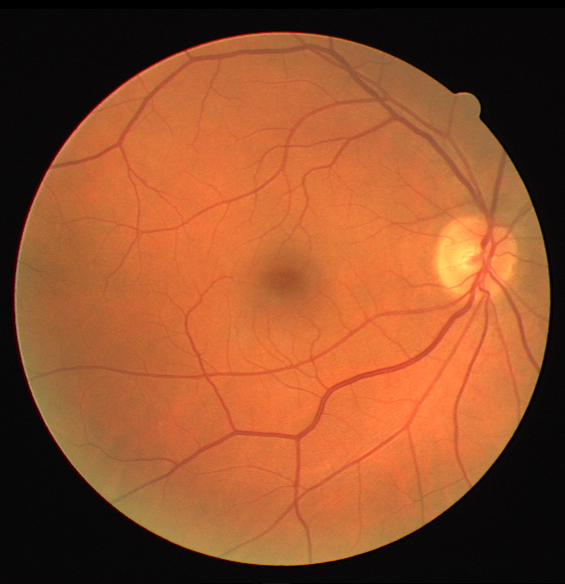}
\caption{DRIVE file 16\_test.}
\label{fig_1}
\end{figure}

\begin{figure}[H]\centering
\includegraphics[width=3.2in]{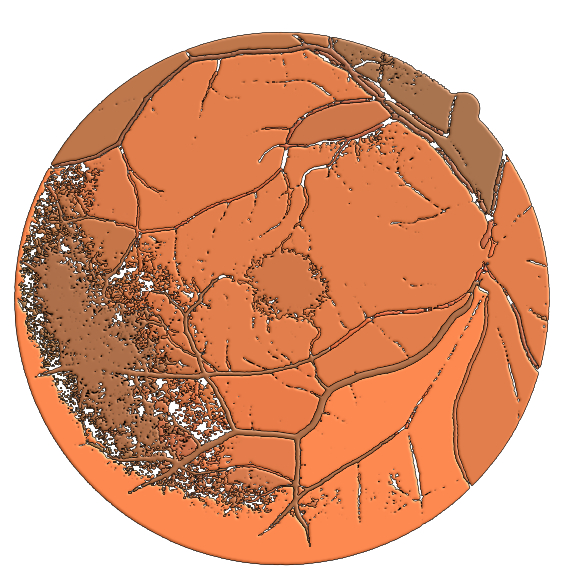}
\caption{Level 1 of the segmentation hierarchy. Each segment is a connected open set in image plane, with simple closed curves as boundary components.  Each boundary component is a level set of the bilinear interpolation of the image green channel. These 1,515 segments are pairwise disjoint. Most of the segments are too small to see without magnification. The largest vessels are represented by a single or only a few segments.}
\label{fig_2}
\end{figure}

\begin{figure}[H]\centering
\includegraphics[width=3.2in]{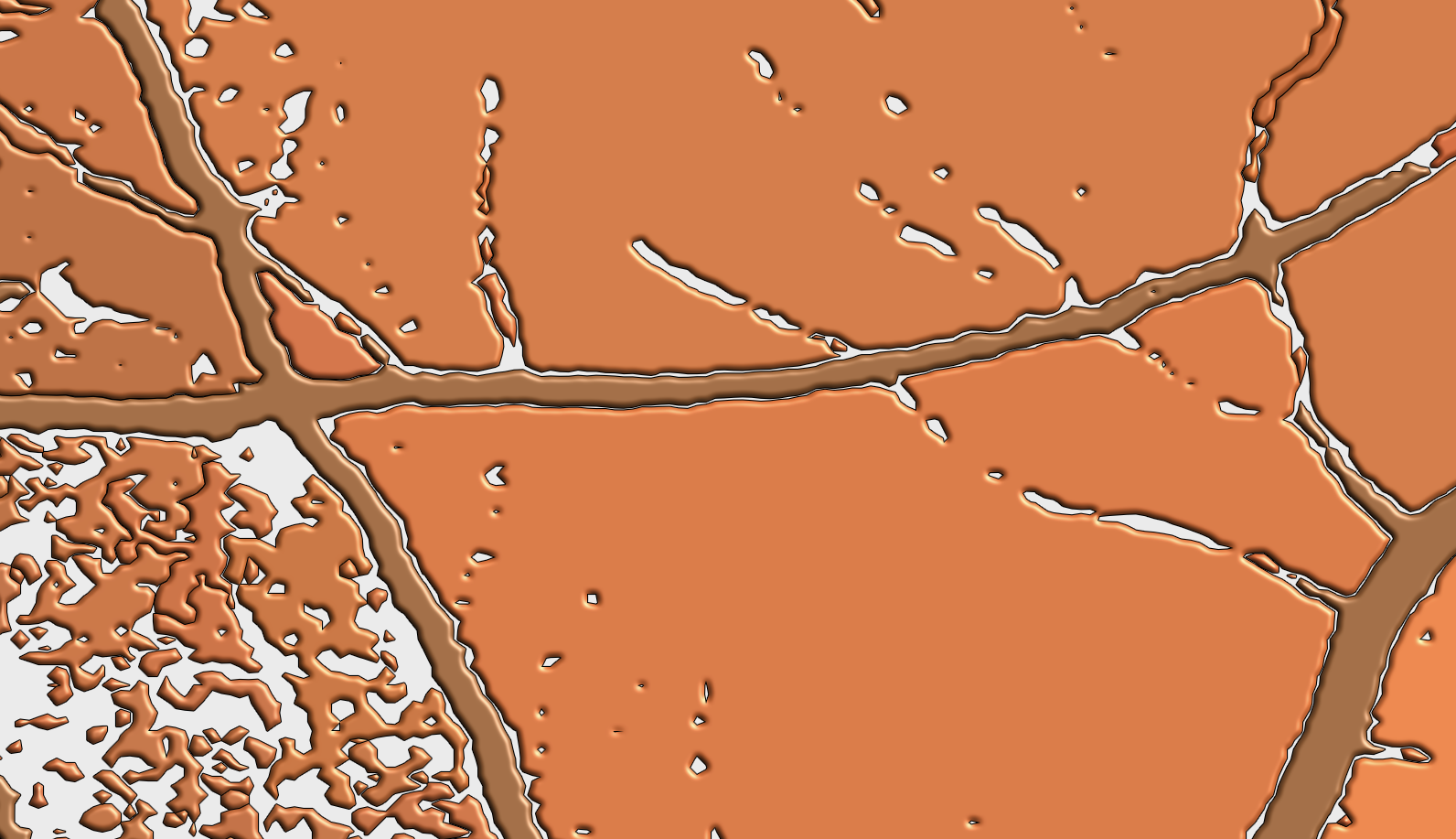}
\caption{Magnification of figure \ref{fig_2} showing single large vessel segment in context, as well as segment holes that also indicate vessel edges.  Analytic and geometric analysis may identify vessel edges by their straight, narrow, parallel characteristics.}
\label{fig_3}
\end{figure}

\begin{figure}[H]\centering
\includegraphics[width=3.3in]{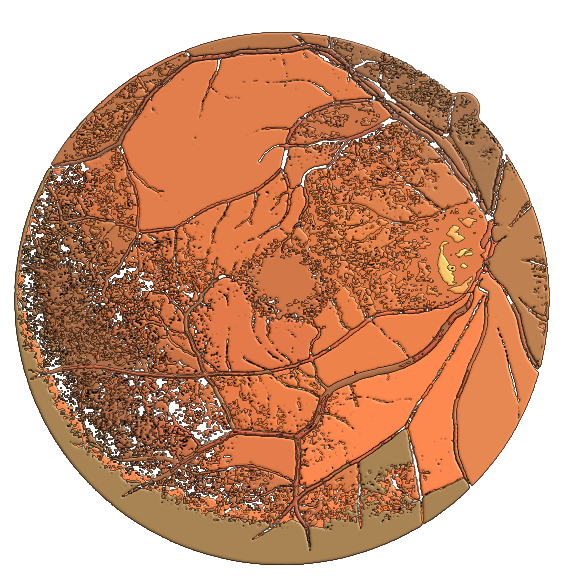}
\caption{Segmentation layer 2 nested within segments of layer 1 (figure \ref{fig_2}). The additional 3,445 segments include new vessel segments lying within individual layer 1 segments, as also provide supporting evidence for layer 1 vessel edges when the layers have coincident long, straight, parallel boundaries.}
\label{fig_4}
\end{figure}

\begin{figure}[H]\centering
\includegraphics[width=3.3in]{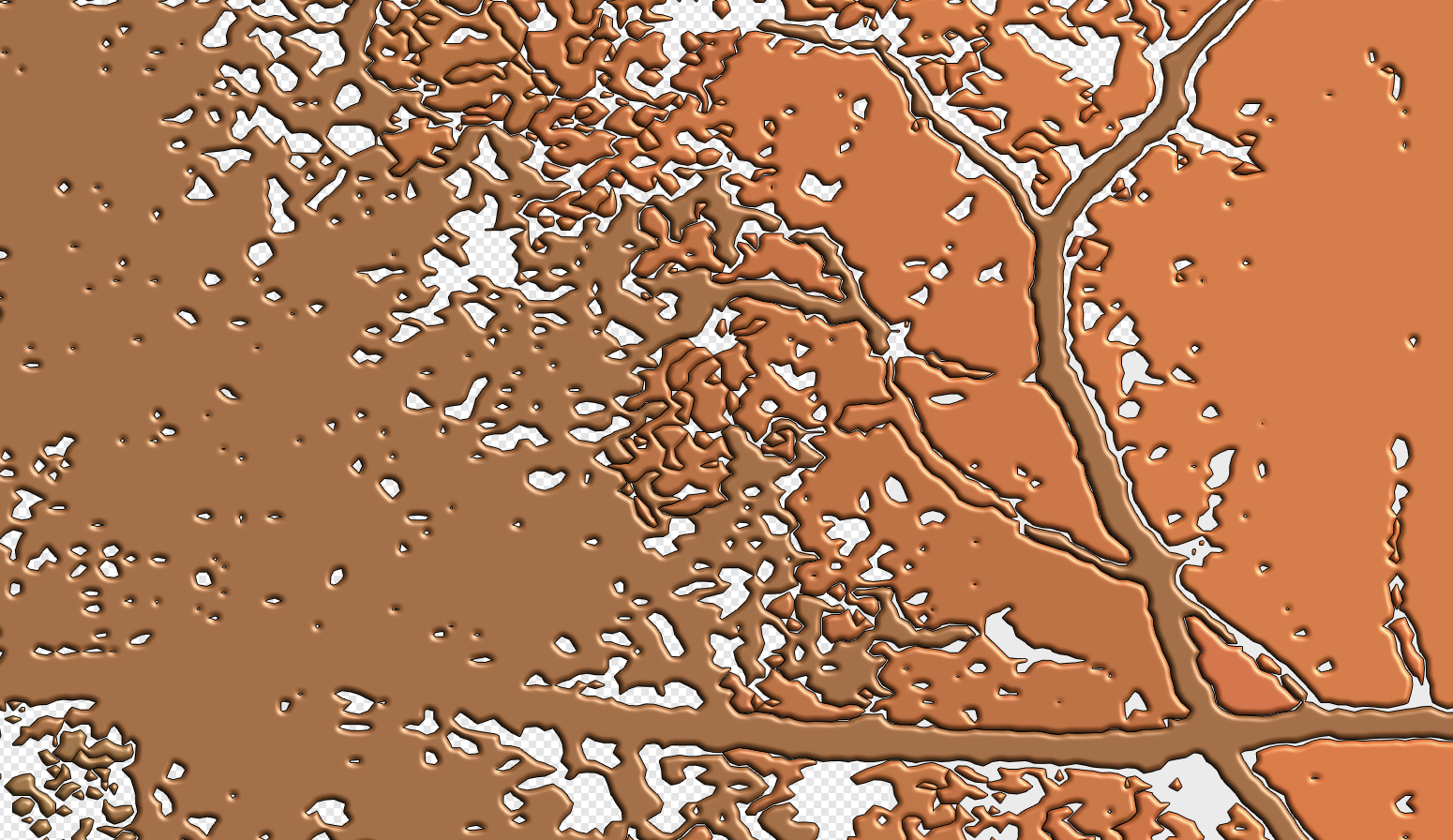}
\includegraphics[width=3.3in]{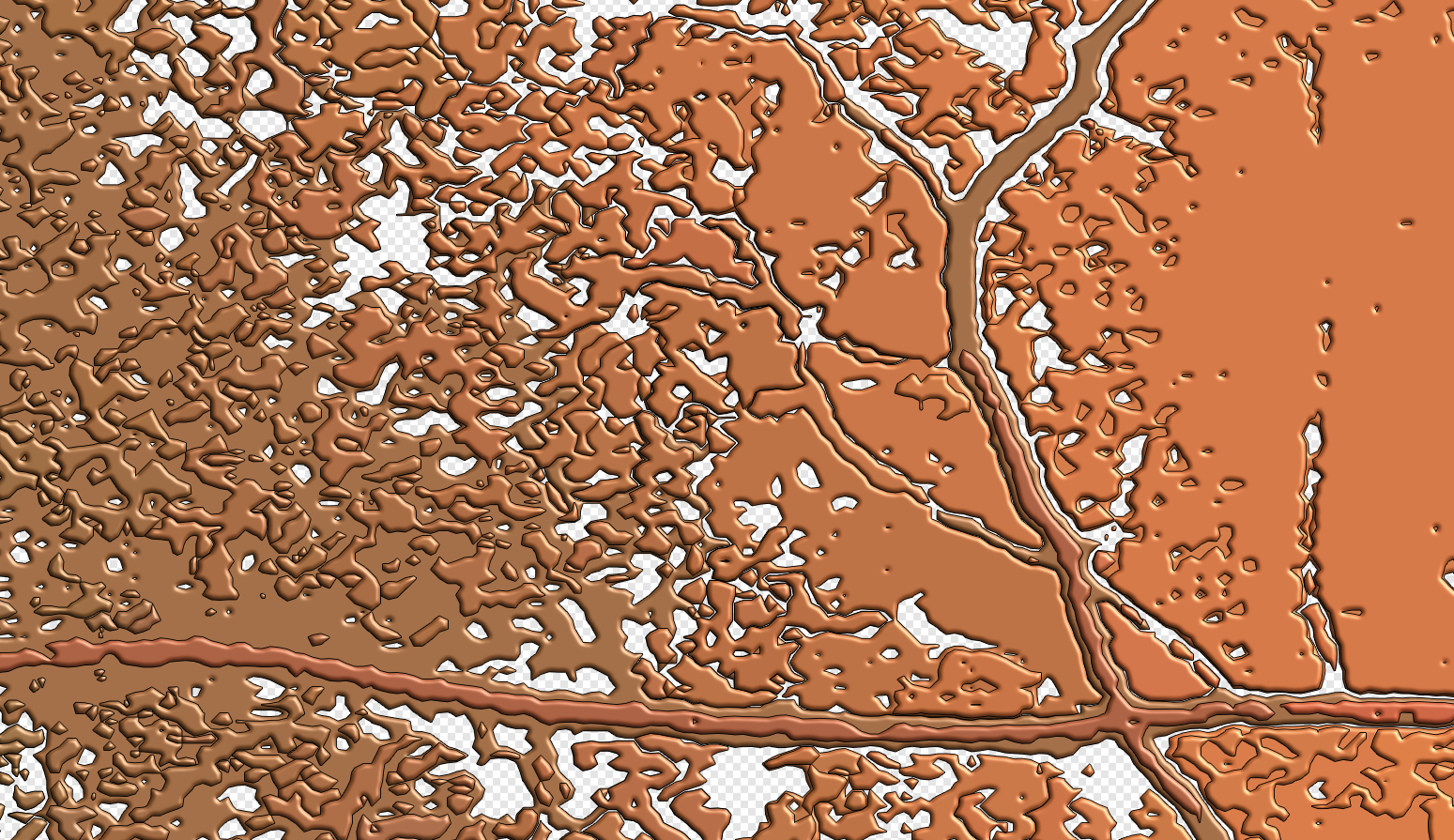}
\caption{Magnification showing how inclusion of segmentation layer 2 (bottom) adds vessel information to layer 1 (top).}
\label{fig_5}
\end{figure}

\section{Topological Retinal Vessel Segmentation Method}

The segmentations of the previous section were generated by computations of varilet image analysis \cite{varilet_image}, as follows:

The green channel of the original image is bilinearly interpolated to a continuous function; from this function we compute the \emph{Reeb graph} \cite{Biasotti}, which in this context has no loops and is also known as the \emph{contour tree} \cite{Carr2000}. The Reeb graph represents the nesting of \emph{level sets} of the interpolated image, with one node for each critical point. The example's Reeb graph has 59,729 nodes.

Working with the nodes and arcs of the Reeb graph as a topological space in its own right (a \emph{graph continuum}), we compute a \emph{persistence lens} -- a hierarchy of closed connected sets (of the graph), having properties related to the pairing of critical points that characterizes \emph{topological persistence} \cite{EdelsbrunnerBook}. The persistence lens is not a unique object; it depends on the parameters that were chosen by analysis of the DRIVE training images. 

Then, from each closed set $S$ of the persistence lens we extract the connected components of $S$'s interior. 
We get the image segments by pulling back each component from the Reeb graph to the image plane.
The boundary of each segment comprises one of more simple closed curves; the circular sequence of points generating the component's SVG shape is calculated from a table created during the initial Reeb graph computation.  These shapes are filled by the SVG even-odd fill rule, thus generating the segmented images.

\section{Results}

The segmentation parameters that resulted from experience with the DRIVE training data were intentionally kept as simple as possible because of the small sample size (20 each, train \& test). 

The segmentation parameters constitute a discrete set of technical alternatives, together with a scale space selection of one or more SVG segmentation images.  
The training experience resulted in a technical parameter set, denoted $P$, and delivery of two images: the segmentation hierarchies of depths one and two\footnote{Additionally, each image was topologically filtered by removing all lens regions having area less that ten square pixels.}. 

Please note that all evaluations are subjective, indicating the \emph{potential} for further automation that would be objectively evaluated.

Methodically running the DRIVE test data through the segmentation engine, parameter set $P$ and two-level scale space selection led to subjectively satisfactory results on 12 out of 20 test images, with partially satisfactory results on 3 of 20\footnote{Images 07\_test, 11\_test and 13\_test provide little information at segmentation depth one.}, and no results for 5 of 20 due to unrecoverable software errors\footnote{Open source libraries used by the author threw errors related to image file handling; the errors are not related to varilet image processing.}.
These images are available at \url{http://varilets.org}.

\section{Comparison with Selected Retinal Vessel Segmentation Methods}

Use of \emph{computational topology} is new for retinal vessel segmentation, but related techniques have been used with brain images \cite{Pepe2012, Segonne, Shi, Chung}, cephalometry \cite{Makram},  liver CT scans \cite{Adcock}, endoscopy images \cite{Dunaeva} and vascular networks  \cite{bendich2016, Cha}.
Vessel branching topology has been used for retinal image classification \cite{Zhu}.

Varilet image analysis is based in part on \emph{level sets}, also used by other segmentation methods \cite{QianZhao, Gongt, Dizdaroglu2014, Lathen}.
Varilet analysis is distinguished from these methods by exploitation of level set relationships as expressed in the Reeb graph, and also by restriction to level sets which are simple closed curves (which is not generally the case).

The present method provides two levels of segmentation from which vessel information may be extracted); these segmentations derive from a larger \emph{scale space}  of \emph{image simplifications} \cite{varilet_image}. Scale space has been used for retinal vessel segmentation \cite{Martinez-PŽrez1999}, as have image simplifications \cite{Sreejini2015253} and multiresolution analysis \cite{Cai2006, Wang, Wang2007}.

Many of the referenced methods recognize vessels by their parallel edges. Similarly, for segments generated by the present method, portions of segment boundaries may be assigned probabilities of being a vessel edge.  Analytic methods iterating through chained segments may probabilistically identify linear and parallel boundary portions. In figures \ref{fig_2} \& \ref{fig_3} we see that such boundary portions may belong to a segment representing the vessel itself, or may belong to a segment that surrounds a vessel.

\section{Integration with Pixel-Based Vessel Segmentation Methods}
\label{int_sec}

\begin{figure}[H]\centering
\includegraphics[width=3.3in]{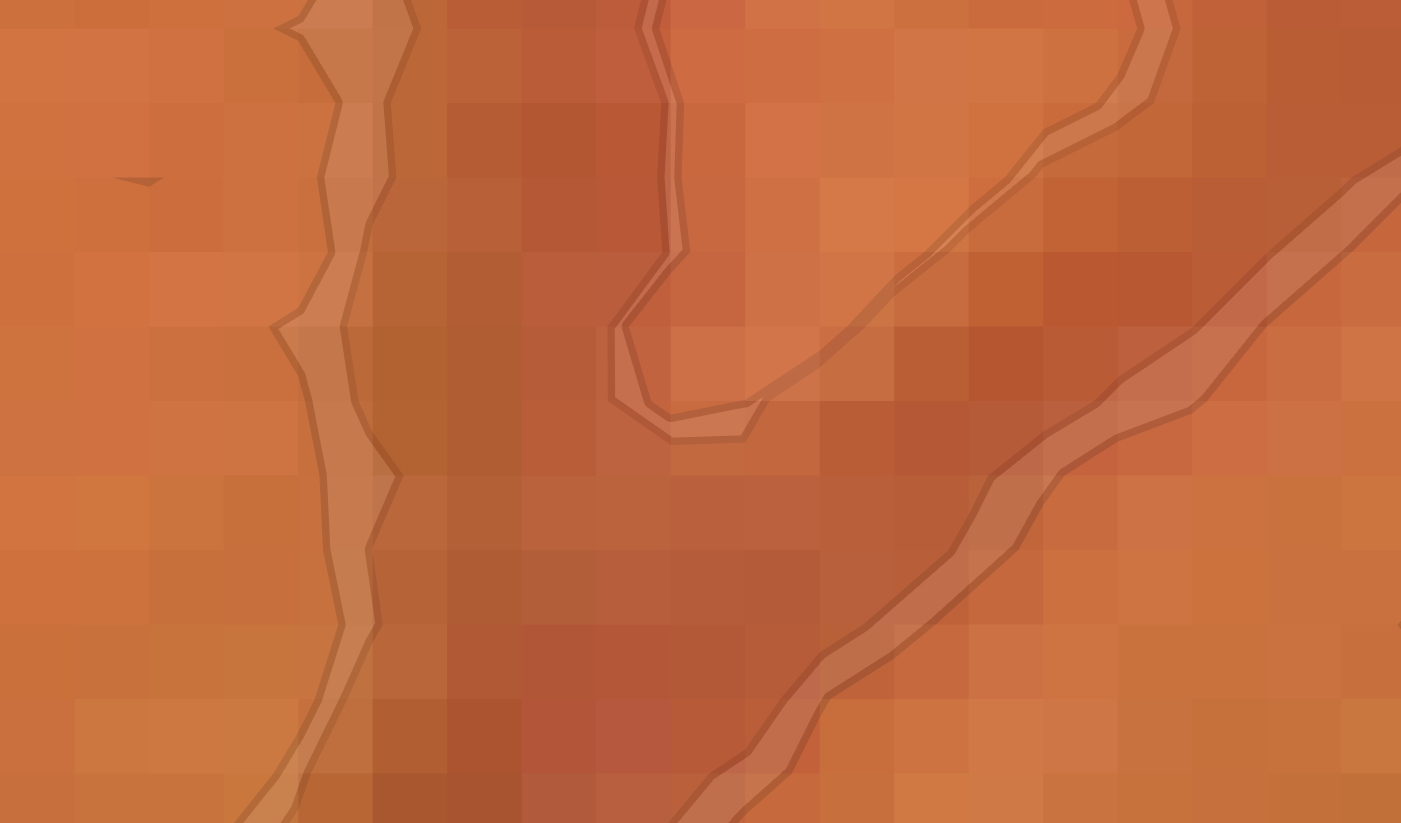}
\caption{Vessel segment in  registration with image pixels.}
\label{fig_6}
\end{figure}

Many of the referenced methods use probabilities to measure the likelihood that each image pixel is part of a vessel; we now describe how these methods might be used in combination with the present method. Suppose we have assigned a probability that a segment edge is a vessel edge; we may then combine that probability with the probabilities of the pixels underlying the segment. Refer to text of figures \ref{fig_3} \& \ref{fig_4}, and see figure \ref{fig_6}.

\section{Conclusion}

We have applied varilet image analysis to retinal images from the DRIVE database, demonstrating a high degree of vessel capture by a two-level hierarchical topological segmentation represented as scalable vector graphics. 

We hope this short paper will interest researchers in application of topological segmentation within imaging and diagnostic tools.

\bibliographystyle{eg-alpha}
\bibliography{../../../varilets}

\end{document}